\documentclass[preprint, showpacs, preprintnumbers, amsmath,amssymb]{revtex4}

\usepackage{graphicx}
\usepackage{dcolumn}
\usepackage{tensor}
\usepackage{fontenc}

\begin{document}

\title{ The Klein-Gordon equation in Machian model }
\author{Bin Liu}
\author{Yun-Chuan Dai}
\author{Xian-Ru Hu}
\author{Jian-Bo Deng}
\email{dengjb@lzu.edu.cn} \affiliation{
Institute of Theoretical Physics, Lanzhou University \\
Lanzhou 730000, People's Republic of China}

\date{\today}

\begin{abstract}

The non-local Machian model is regarded as an alternative theory of gravitation which states that all particles 
in the Universe as a 'gravitationally entangled' statistical ensemble. It is shown that the Klein-Gordon equation 
can be derived within this Machian model of the universe. The crucial point of the derivation is 
the activity of the Machian energy background field which causing a fluctuation about the average momentum of a particle, 
the non-locality problem in quantum theory is addressed in this framework.


\end{abstract}
\maketitle

\section{introduction}

It has been demonstrated during last decade that a wide class of field theoretical models can be given a thermodynamic interpretation.
For instance, both the Newton law and the Einstein equation can be derived by using thermodynamical methods \cite{f1}\cite{f2}\cite{f3}. 
Similar investigations showed that Coulomb law and the Maxwell equations could also be derived in such way \cite{f4}. 
Furthermore, the analogies between classical statistical mechanics and quantum mechanics are also well focused \cite{f5}\cite{f6}\cite{f7}.\\
\indent Since the conception that classical field theory explained by the laws of thermodynamics and statistical physics generates many questions, 
those surprising relationships are still regarded as just analogies. 
The central problem is the necessity of modifying the notion of locality for the 
underlining theories when considering physical theories as emergent phenomena from some thermodynamical processes. 
It is known that at least quantum mechanic models with hidden-variables lead to the problems of locality \cite{f8}. 
In fact, the research for an acceptable model of quantum gravity already has motivated 
several authors to introduce non-local corrections for special \cite{f9} and general \cite{f10} relativity theories.\\
\indent One example of thermodynamical approach is Machian model of gravity \cite{f11}. 
The new feature of this model is the non-locality which assumes that all particles of the universe 
are involved in the gravitational interaction between any two particles. 
Adopted the thermodynamical-statistical formulation, 
the rest mass of a particle is a measure of its long-range collective gravitational interactions with all other particles 
inside the horizon \cite{f12}. In ref.\cite{f13}, the Machian model imitates basic features of special and general relativity theories. 
Ref.\cite{f14} has shown that both the Schr\"odinger equation and the Planck constant can be derived within this Machian model of the universe.\\
\indent Non-locality is known to all as an essential feature of quantum mechanics \cite{f15}. 
It makes sense to suppose that the non-locality in the Machian model of gravity and quantum mechanics have the same origin. 
Many efforts have been made to derive the equations from thermodynamical-statistical approach. Some of attempts are like: 
using the nonequilibrium thermodynamics \cite{f16}, within the hydrodynamical interpretation of quantum mechanics \cite{f17}, 
in Nelson's stochastic mechanics \cite{f18}, from the 'exact uncertainty principle' \cite{f19}, 
or using the hidden-variable theory of de Broglie and Bohm \cite{f20}  etc.\\
\indent  In this paper, by using classical statistical mechanics for the Machian universe, 
the Klein-Gordon Equation is obtained. In fact, the equation can be directly derived 
by means of action from the non-local Machian energy of a particle. 
The only inputs would be given by the existence of an associated space-pervading field with the non-local Machian energy, 
i.e. next to the usual action function describing a classical physical system, interaction from all particles of the universe is considered. 
It is thus assumed that the energy of the non-local gravitational interactions of the particle pervades all of universe. 
Due to the notion of the Machian energy background field, the classical Lagrangian should be modified, 
i.e. there should be a fluctuation term concerning the average momentum of a particle. 
This is the key point of deriving the Klein-Gordon equation from the variation principle as shall be shown.\\
\indent The structure of this paper is as following. 
First, we review the idea of Machian model. 
Then the energy balance equation is given and the standard relativistic formulas are 
proposed by adopting Mach's interpretation of inertia \cite{f12}\cite{f13}. 
Second, due to the assumed activity of the Machian energy background field, the complete action integral of a particle is obtained. 
Third, from the formalism of classical statistical mechanics, 
we derive the equation of motion of particle and show that it is equivalent to the Klein-Gordon equation. 
Finally, the paper ends with a brief summary.\\

\section{THE MODEL OF MACHIAN UNIVERSE}\label{SecB}

In Machian model all the particles in the Universe are non-locally interacting to each other 
and the universe can be considered as a statistical ensemble of 'gravitationally entangled' particles \cite{f12}\cite{f13}\cite{f14}. 
When considering the gravitational interaction of the close objects, 
one can replace the distant universe by a spherical shell of the effective mass $M$ and the effective radius $R$ 
which is consisting of the ensemble of $N$ uniformly distributed identical particles of gravitational mass $m_g$. 
One of the main assumptions of the model is that the 'universal' gravitational potential of any particle from the ensemble is \cite{f21}:
\begin{equation}\label{eq1}
\phi=-\frac{MG}{R}\approx c^2
\end{equation}
\indent The non-local Machian interactions between particles reveal the main differences from the standard physics, 
for each particle in the 'gravitationally entangled' universe interacts with all the other $(N-1)$ particles. 
So the number of interacting pairs is $N(N-1)/2$, and the Machian energy of a single particle take the form of \cite{f12}:
\begin{equation}\label{eq2}
E_0=-\frac{N(N-1)}{2}\frac{2Gm^2_g}{R}\approx -\frac{GN^2m^2_g}{R}=-m_g\phi
\end{equation}
It is shown that the assumption, the whole Universe is involved in local interactions, 
weakens the observed strength of gravity effectively by a factor related to the number of particles in the Universe.\\
\indent Adopting the energy additivity as in the Newtonian realm the energy balance equation for a particle is given as:
\begin{equation}\label{eq3}
E=E_0+T+U
\end{equation}
Here $T$ is its kinetic energy with respect to the preferred frame 
and $U$ corresponds to all local interactions causing an acceleration of the particle.\\
\indent Aspects of relativity can be explained in this model with a preferred frame using Mach¡¯s principle \cite{f13}. 
Because having any kinetic energy $T$, the value of $E_0$ should be the same for all inertial observers $(U=0)$ 
in the homogeneous universe, the relativity principle recovers:
\begin{equation}\label{eq4}
m_0=m\sqrt{1-\frac{v^2}{c^2}}
\end{equation}
Here $m_0$ is the inertial mass and the definition of inertia in this case reduces to the equivalence principle $m_0=m_g$. 
It is shown that the mass transformation has been found and the well known 'relativistic' effect of the particle's 
inertia growth depending on its velocity has a dynamical nature which is a consequence of the energy balance condition (\ref{eq3}). 
All other 'relativistic' effect could be obtained as well, including the formula of the linear momentum:
\begin{equation}\label{eq5}
p_{\mu}p^{\mu}=-m^2c^2
\end{equation}
So the formulas of special relativity and the justification of the 4-dimensional geometrical formulation appear in this framework naturally.

\section{THE COMPLETE ACTION INTEGRAL OF A PARTICLE}\label{SecC} 
In this section base on the assumed activity of the Machian energy background field, 
an additional fluctuation which is subject to the average momentum of a particle is produced 
and the complete action integral is obtained.\\ 
\indent In ref.\cite{f14} the description of physical systems by ensembles 
may be introduced at quite a fundamental level using notations of probability and action principle. 
We start with a particle:
\begin{equation}\label{eq6}
A=-\int^{t_2}_{t_1}Edt
\end{equation}
\indent Furthermore the portion of the action of the universe for a single member of the ensemble should 
take the same order of the Planck constant $(A\sim -2\pi\hbar)$.\\
\indent As in the models \cite{f16}\cite{f17}\cite{f18}\cite{f19}\cite{f20}, 
we assume that the probability of finding the particle in the configuration space is described by a probability density $P(x,t)$. 
It obeys the normalization condition:
\begin{equation}\label{eq7}
\int Pd^3x=1
\end{equation}
\indent By means of $P(x,t)$ the action integral (\ref{eq6}) for a particle in the ensemble can be written as:
\begin{equation}\label{eq8}
A=-\int PE d^3xdt
\end{equation}
\indent Now consider the case when only one 'free' particle moves with respect to the 
preferred frame and local potential acting on the particle would not be considered $(U=0)$. 
We introduce the action function $S(x,t)$ \cite{f14}
\begin{equation}\label{eq9}
S=p_{\mu}x^{\mu}+const
\end{equation}
Here $S(x,t)$ is related to the particle velocity $v(x,t)$ and momentum $p_{\mu}$ via:
\begin{equation}\label{eq10}
v_{\mu}=\frac{\partial_{\mu} S}{m}~,~~~~~~~p_{\mu}=\partial_{\mu} S
\end{equation}
\indent The action integral (\ref{eq8}) in this case has the form:
\begin{equation}\label{eq11}
A=\int P(\frac{\partial S}{\partial t}+\frac{\partial_{\mu}S\partial^{\mu}S}{m})dtd^3x
\end{equation}
\indent For the case when all $N$ particles in the universe are at rest with respect to each other, we get:
\begin{equation}\label{eq12}
S_0=-Et+const
\end{equation}
Where the particle's energy (\ref{eq3}) coincides with the energy of its non-local gravitational interactions 
with the universe and takes the form as $E=E_0=-m_g\phi$. In other words action function $S_0$ means the 
activity of the Machian energy background field. We shall assume in the following that this energy would cause 
an additional velocity fluctuation field relative to the original state of motion of a particle 
\begin{equation}\label{eq13}
u_{\mu}=\frac{\partial_{\mu} S_0}{m}
\end{equation}
\indent Due to the assumed Machian energy which contributes a non-classical dynamics, $p_{\mu}=\partial_{\mu} S$ 
is only an average momentum with momentum fluctuations $f_{\mu}$ around $\partial_{\mu} S$. 
Thus, the physical momentum is:
\begin{equation}\label{eq14}
p_{\mu}=\partial_{\mu} S+f_{\mu}
\end{equation}
No particular underlying physical models tend to be assumed for the fundamentally nonanalyzable \cite{f20}. 
Instead, one could regard $'f_{\mu}'$ as the fluctuations. 
We could retain the notion of particle trajectories and consider a physically motivated proposal of the form (\ref{eq14}). 
It is natural to assume that the momentum fluctuations $f_{\mu}$ are linearly uncorrelated with the average momentum $p_{\mu}=\partial_{\mu} S$.
Hence:
\begin{equation}\label{eq15}
\int P(\partial_{\mu} S \cdot f_{\mu})d^3x=0
\end{equation}
Thus, with $p$ given by equation (\ref{eq14}), 
it is proposed that the complete action integral of a particle immersed in the Machian energy background field is given by: 
\begin{equation}\label{eq16}
A=\int P(\frac{\partial S}{\partial t}+\frac{\partial_{\mu}S\partial^{\mu}S}{m})dtd^3x+\frac{1}{m}\int(\Delta f)^2dt
\end{equation}
Where $(\Delta f)^2$ is the average of square momentum fluctuation given by:
\begin{equation}\label{eq17}
(\Delta f)^2=\int Pf_{\mu} \cdot f_{\mu}d^3x
\end{equation}
Base on the above discussion which states that the additional momentum fluctuation $f_{\mu}$ is due to 
the activity of the Machian energy background field. In this case $f_{\mu}$ could be described by the action function $S_0$ in eq.(\ref{eq12})
\begin{equation}\label{eq18}
(\Delta f)^2=\int P\partial_{\mu}S_0\partial^{\mu}S_0d^3x
\end{equation}
Combining eqs.(\ref{eq16}) and (\ref{eq18}), we get:
\begin{equation}\label{eq19}
A=\int P(\frac{\partial S}{\partial t}+\frac{\partial_{\mu}S\partial^{\mu}S}{m}+\frac{\partial_{\mu}S_0 \partial^{\mu}S_0}{m})dtd^3x
\end{equation}
This is the action integral for a 'free' particle involving a fluctuation field in a four-dimensional volume 
in accordance with relativistic kinematics. 

\section{DERIVATION OF KLEIN-GORDON EQUATION}\label{SecD}
The formula (\ref{eq19}) is written for a particle moving with respect to the Universe. Moreover, the Lagrangian becomes:
\begin{equation}\label{eq20}
\mathcal {L}=P(\frac{\partial S}{\partial t}+\frac{\partial_{\mu}S\partial^{\mu}S}{m}+\frac{\partial_{\mu}S_0 \partial^{\mu}S_0}{m})
\end{equation}
\indent Upon the principle of least action, we set $\delta P=\delta S=0$ at the boundaries. 
First we perform the fixed end-point variation in $S$ of the Euler-Lagrange equation
\begin{equation}\label{eq21}
\frac{\partial{\mathcal {L}}}{\partial S}-\partial_{\mu}\left[ \frac{\partial{\mathcal {L}}}{\partial (\partial_{\mu}S)}\right]=0
\end{equation}
then it provides the covariant continuity equation:
\begin{equation}\label{eq22}
\frac{\partial P}{\partial t}+\frac{2\partial_{\mu}(P\partial^{\mu}S)}{m}=0
\end{equation}
\indent In the simplest case when a typical particle of the 'universal' ensemble is at rest with respect to the Universe, we have:
\begin{equation}\label{eq23}
\frac{\partial P_0}{\partial t}=0
\end{equation}
Since $P_0$ is time-independent, this equation shows that all the other $(N-1)$ particles of the ensemble 
are at rest with respect to the preferred frame. In another case when only one particle moves but all other $(N-1)$ particles are at rest, 
we have a stationary distribution function as well. Hence, the continuity equation (\ref{eq22}) reduces to:
\begin{equation}\label{eq24}
\partial_{\mu}(P\partial^{\mu}S)=0
\end{equation}
Motivated by ref. \cite{f16}, the solution of the above equation can be written in the form:
\begin{equation}\label{eq25}
P=P_0e^{\frac{2S_0}{\hbar}}
\end{equation}
As to this probability density, we note that the only movement of the particle deviating from the classical path 
must be due to the activity of the Machian energy background as given in eqs.(\ref{eq12}) and (\ref{eq18}), respectively. 
Moreover we obtain from equation (\ref{eq25}) that:
\begin{equation}\label{eq26}
\frac{\partial_{\mu}P}{P}=\frac{2}{\hbar}\partial_{\mu}S_0
\end{equation}
In terms of the definition of the velocity fluctuation field (\ref{eq13}), we have:
\begin{equation}\label{eq27}
u_{\mu}=\frac{\partial_{\mu}S_0}{m}=\frac{\hbar}{2m}\frac{\partial_{\mu}P}{P}
\end{equation}
where this relation could lead to the momentum fluctuation 
\begin{equation}\label{eq28}
mu_{\mu}=\frac{\hbar}{2}\frac{\partial_{\mu}P}{P}
\end{equation}
\indent Now performing the fixed end-point variation in $P$ of the Euler-Lagrange equations
\begin{equation}\label{eq29}
\frac{\partial{\mathcal {L}}}{\partial P}-\partial_{\mu}\left[ \frac{\partial{\mathcal {L}}}{\partial (\partial_{\mu}P)}\right]=0
\end{equation}
then with eq.(\ref{eq27}) it provides the Hamilton-Jacobi equation 
\begin{equation}\label{eq30}
\frac{\partial S}{\partial t}+\frac{\partial_{\mu}S\partial^{\mu}S}{m}-mu_{\mu}u^{\mu}-\hbar \partial_{\mu}u^{\mu}=0
\end{equation}
Because:
\begin{equation}\label{eq31}
\frac{\partial S}{\partial t}=-E=-mc^2
\end{equation}
we get:
\begin{equation}\label{eq32}
\partial_{\mu}S\partial^{\mu}S-m^2c^2-m^2u_{\mu}u^{\mu}-m\hbar \partial_{\mu}u^{\mu}=0
\end{equation}
As the last two terms of LHS are identical to the relativistic expression 
for the "quantum potential" term \cite{f19}. 
This potential is usually interpreted as the source of uncertainty and non-locality, 
which make quantum systems different from classical ones described by pure Hamilton-Jacobi equation.
However, the motion of 'gravitationally entangled' particles are at any distance in this approach.
Then we have obtained the relativistic Hamilton-Jacobi-Bohm equation
\begin{equation}\label{eq33}
\partial_{\mu}S\partial^{\mu}S-m^2c^2-\hbar^2\frac{\square \sqrt{P}}{\sqrt{P}}=0
\end{equation}
\indent It is known that a justification of the quantization condition is essential to derive the quantum 
equations using concepts from classical physics \cite{f16}\cite{f17}\cite{f18}\cite{f19}\cite{f20}. 
In our approach we use the Madelung transformation \cite{f17}
\begin{equation}\label{eq34}
\psi=\sqrt{P}e^{-\frac{iS}{\hbar}}
\end{equation}
\indent As is well known, the eqs.(\ref{eq24}) and (\ref{eq33}), 
together with the introduction of the above complex-numbered 'wave function' (\ref{eq34}), we obtain the usual Klein-Gordon equation
\begin{equation}\label{eq35}
\left(\square +\frac{m^2c^2}{\hbar^2}\right)\psi=0
\end{equation}
We have thus succeeded in deriving the Klein-Gordon equation from the action with the assumptions relating to 
the Machian energy background field associated to each particle.

\section{conclusion}\label{SecE}
In this paper the Machian model which is an alternative theory of gravitation has been introduced 
and it could lead to the main features of the special relativity. Then a quantum behavior of matter 
in the Machian model of the universe is considered. 
The notion of non-locality that all particles in the universe interact with each other via a long-range gravitational force allows us to treat all the particles in the universe as the members of a 'universal' statistical ensemble. 
Using classical statistical mechanics, the Klein-Gordon equation is derived from the system of non-linear continuity 
and Hamilton-Jacobi equations, 
as well as the relativity principle in the Machian model with the preferred frame. 
The crucial point of the derivation is the activity of the Machian energy background field 
which causing a fluctuation about the average momentum of a particle, 
the non-locality problem in quantum theory is addressed in this framework.

\end{document}